\documentclass[twocolumn,aps, prl, groupedaddress,epsf]{revtex4}

\usepackage{graphicx}

\begin{document}

\title{Thermoelectric and Magnetothermoelectric Transport Measurements of Graphene}

\author{Yuri M. Zuev$^1$}
\author{Willy Chang$^2$}
\author{Philip Kim$^2$}

\affiliation{Department of Applied Physics and Applied Mathematics$^1$ and Department of
Physics$^2$, Columbia University, New York, New York, 10027, USA}

\date{\today}

\begin{abstract}
The conductance and thermoelectric power (TEP) of graphene is simultaneously measured using microfabricated heater and thermometer electrodes. The sign of the TEP changes across the charge neutrality point as the majority carrier density switches from electron to hole. The gate dependent conductance and TEP exhibit a quantitative agreement with the semiclassical Mott relation. In the quantum Hall regime at high magnetic field, quantized thermopower and Nernst signals are observed and are also in agreement with the generalized Mott relation, except for strong deviations near the charge neutrality point.
\end{abstract}

\maketitle

Recent research on graphene, a zero-mass, zero-gap, ambipolar material system, has focused on its unique electronic properties~\cite{geim_review}. Although there have been theoretical investigations of thermal and thermoelectric properties of this 2-dimensional material~\cite{lofwander_fog,sta_per,per_santos,dora_thal,foster_largeN,foster_imbalance,sachdev_int}, only an indirect measurement of the thermal conductivity~\cite{lau_thermal} employing Raman spectroscopy has been reported recently, and thermoelectric transport has yet to be explored experimentally. In general, the thermoelectric power (TEP) is of great interest in understanding electronic transport due to its extreme sensitivity to the particle-hole asymmetry of a system~\cite{Mott, TEP_macdonald}. Measurement of the thermoelectric properties of graphene can thus elucidate details of the electronic structure of the ambipolar nature of graphene that cannot be probed by conductance measurements alone. Furthermore, employing the semiclassical Mott relation~\cite{Mott}, one can verify the validity of the Boltzmann approach to transport in graphene by comparing the simultaneously measured conductance and TEP at different chemical potentials~\cite{jsmall_prl}. The application of a magnetic field in addition to a thermal gradient provides a valuable experimental tool to investigate magnetothermoelectric effects, such as the Nernst effect  in 2-dimensional (2D) electronic systems in the quantum Hall (QH) regime~\cite{jon_girv,oji,obl_klitz,streda}.

In this letter, we first discuss the temperature and carrier density dependence of the TEP measured in mesoscopic graphene samples, and its relation with the Mott formula. We then discuss the magnetothermopower measurement where we observe unusual quantized TEP in the presence of the QH effect.

\begin{figure}
\includegraphics[width=1.0\linewidth]{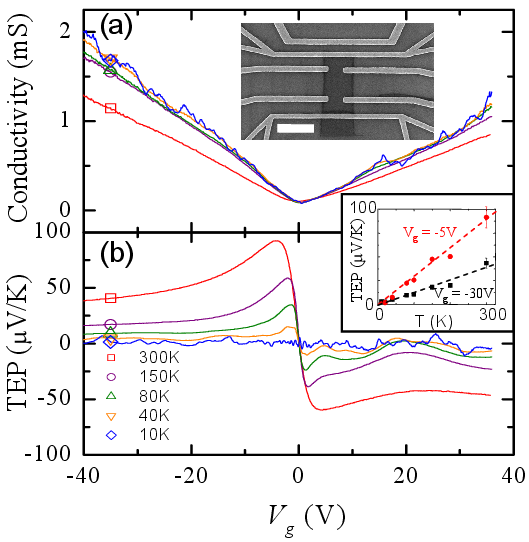}
\caption{(a) Conductivity and  (b) TEP  of a graphene sample as a function of $V_g$ for $T=300$~K (square), 150~K (circle), 80~K (up triangle), 40~K (down triangle), and 10~K (diamond). (Upper Inset) SEM image of a typical device, the scale bar is $2\mu$m. (Lower Inset) TEP values taken at $V_{g}=-30$~V (circle) and -5~V (square). Dashed lines are linear fits to the data.}\label{fig1}
\end{figure}

Graphene samples are fabricated using the mechanical exfoliation method on 300~nm SiO$_{2}$/Si substrate. The graphene carrier density is adjusted with applied gate voltage to the degenerately doped Si substrate. Electrodes are defined with standard electron beam lithography, followed by Ti/Au (3/30~nm) evaporation and liftoff processes. The TEP measurement technique was described in detail elsewhere~\cite{jsmall_prl}. In brief, a controlled temperature gradient, $\Delta T$, is applied to the sample by a microfabricated heater while the resulting thermovoltage, $\Delta V$, is measured by the voltage probes to acquire the TEP, $S=-\frac{\Delta V}{\Delta T}$ (see the upper inset of Fig.~1 for device layout). Local temperature variations are measured with two metal 4-probe micro-thermometers. Oxygen plasma etching is used to isolate the graphene flake electrically from the heater electrode. We use Raman spectroscopy and QH measurements in order to verify that our samples are single layer graphene ~\cite{geim_QHE,zhang_kim}. Devices are annealed in a vacuum cryostat to 400~K in order to improve mobility and electron-hole symmetry. We measure both the 2-terminal mesoscopic conductance, $G$, utilizing the outer electrodes and the 4-terminal conductivity, $\sigma$, employing the Hall bar geometry. To measure the TEP we apply a low frequency AC signal to the heater electrode and pick up the resulting $2\omega$ thermovoltage reading from the outer electrodes as we change the graphene carrier density with the back gate. To stay in the linear response regime the condition $\Delta T\ll T$ is always satisfied.

In this work, we study 10 different graphene samples with mobility ranging from 1,000 to 7,000 cm$^2$/Vs. Fig.~1a-b show simultaneously measured $\sigma$ and TEP as a function of applied gate voltage $V_g$ for a representative graphene device in the temperature range of 10~-~300~K. As it was shown previously~\cite{geim_review,DasSarma_scatt}, $\sigma\propto V_{g}$ away from the charge neutrality point (CNP). The conductivity becomes minimum at the CNP, corresponding to $V_g=V_D$, where $V_D\approx0$~V for this particular device. The peak value of TEP reaches $\sim80\mu$V/K at room temperature in the majority of devices measured. Notably, the sign of the TEP, which indicates the sign of the majority charge carrier, changes from positive to negative as $V_g$ crosses the CNP. The electron-hole asymmetry observed in the conductance and also in the TEP is most likely due to charged impurities in the oxide substrate~\cite{DasSarma_scatt, Neto_scatt} or the formation of a p-n junction at the electrode contacts~\cite{GG_contact}. Universal conductance fluctuations (UCF) are observed in the conductance for $T\lesssim 30$~K and are quantitatively correlated to the oscillations in the TEP, as we will show later.

While $\sigma$ decreases as $T$ increases, the magnitude of the TEP increases monotonically as $T$ increases. The inset of Fig.~1b shows a linear temperature dependence of TEP at two representative gate voltages, $V_g=-30$~V and $V_g=-5$~V. The linearity of the TEP in $T$ suggests that the mechanism for thermoelectric generation is diffusive thermopower. Phonon-drag components of the thermopower, which usually appear in the form of a non-monotonic temperature dependence of TEP~\cite{mahan_drag}, are not present here, which is consistent with weak electron-phonon coupling in graphene~\cite{bolotin_Tdep,hwang_sarma}.

\begin{figure}
\includegraphics[width=1.0\linewidth]{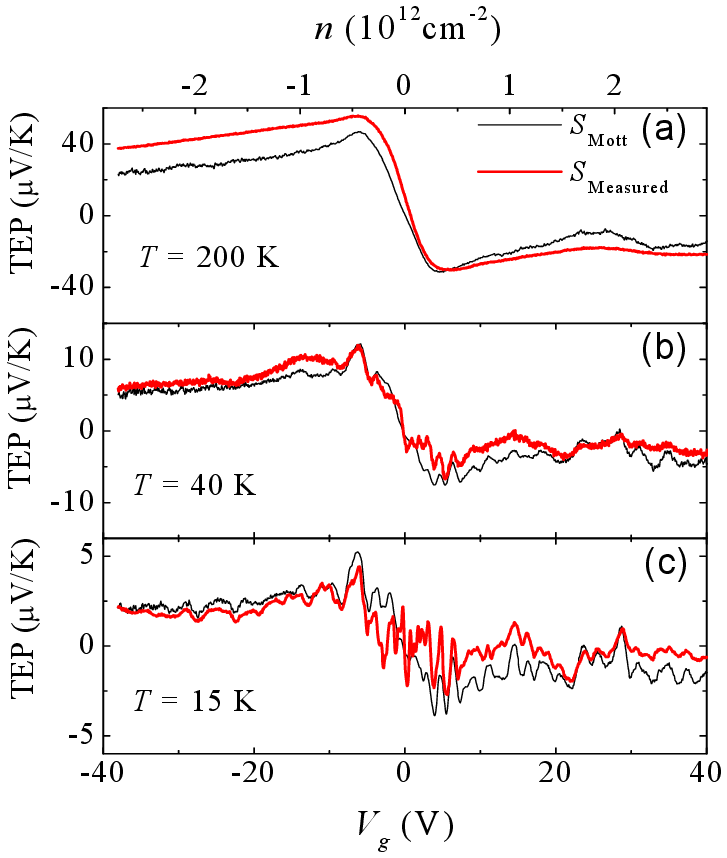}
\caption{ Measured TEP values (thick red) and TEP values calculated with the Mott formula (thin black) using Eq.~ (1) as a function of applied gate voltage for $T=200$~K (a),  40~K (b), and 15~K (c). }
\end{figure}

Within the Boltzmann formulation~\cite{ashcroft_mermin}, the Seebeck coefficient, $S$, equivalent to the TEP, represents the entropy transported per unit charge. This thermodynamic relation provides us with a qualitative understanding of the overall behavior of the observed $S(V_g)$ as $V_g$ tunes the carrier density, $n$, and thus the Fermi energy, $E_F$, of the graphene samples. Far away from the CNP, the electronic system is highly degenerate and the entropy transported in the channel is proportional to the number of thermally activated carriers over the degenerate Fermi sea, leading to $S\sim T/E_F$. Close to the charge neutrality point, however, the electron gas becomes non-degenerate and both electrons and holes are thermally populated. Here the TEP gains opposite contributions from electrons and holes, so $|S|$ decreases as $|n|$ decreases. However, the non-degenerate limit is difficult to achieve in most of the experimental temperature range since the electron-hole puddle formation near the charge neutrality point yields a finite local density $>$10$^{11}$ cm$^{-2}$~\cite{Puddle}, corresponding to $\gtrsim$100~K of energy broadening.

For a quantitative understanding of the TEP data, we now employ the semiclassical Mott relation ~\cite{Mott, TEP_macdonald}. In mesoscopic graphene samples, this relation can be tested by tuning $E_F$ using $V_g$:
\begin{equation}
S=-\frac{\pi^{2}k_{B}^{2}T}{3|e|}\frac{1}{G}\left.\frac{dG}{dV_{g}}\frac{dV_{g}}{dE}\right|_{E=E_{F}}
 \label{Mott1}
 \end{equation}
where $e$ is the electron charge and $k_B$ is the Boltzmann constant. This formula can be evaluated by numerically differentiating the measured $G$ with respect to $V_g$ and considering $\frac{dV_{g}}{dE_{F}}(V_{g})=\frac{\sqrt{|e|}}{C_{g}\pi}\frac{2}{\hbar v_{F}}\sqrt{|\Delta V_{g}|}$, where $\Delta V_{g}=V_{g}-V_{D}$, $\hbar$ is Planck's constant, the Fermi velocity $v_{F}=10^{6}$m/s, and the gate capacitaince $C_{g}=115$~aF$/\mu$m$^{2}$ in our device geometry~\cite{zhang_kim}.  A comparison between the measured TEP and that predicted by the Mott formula is plotted in Fig. 2 for temperatures $T=$15, 40, and 200~K, where we find excellent agreement at lower temperatures. We note that Eq.~(1) is in the form of the mesoscopic formula~\cite{sivan_imry} where the conductance includes the contact resistance~\cite{fnote0}. At higher temperatures, local thermal equilibrium is established, and the mesoscopic form Eq. (1) using $G$ becomes less accurate, as it is already seen in the 200~K data in Fig. 2. Such deviation grows at higher temperatures. Recent theoretical work~\cite{foster_imbalance} predicts that in graphene the enhanced inelastic scattering time of carriers in the presence of electron-electron interactions yields TEP that deviates strongly from the Mott relation. Further experimental work with higher mobility samples (in current samples disorder dominates transport even at room temperature) is needed to elucidate this non-degenerate, high temperature limit of TEP in graphene.

\begin{figure}
\includegraphics[width=1.0\linewidth]{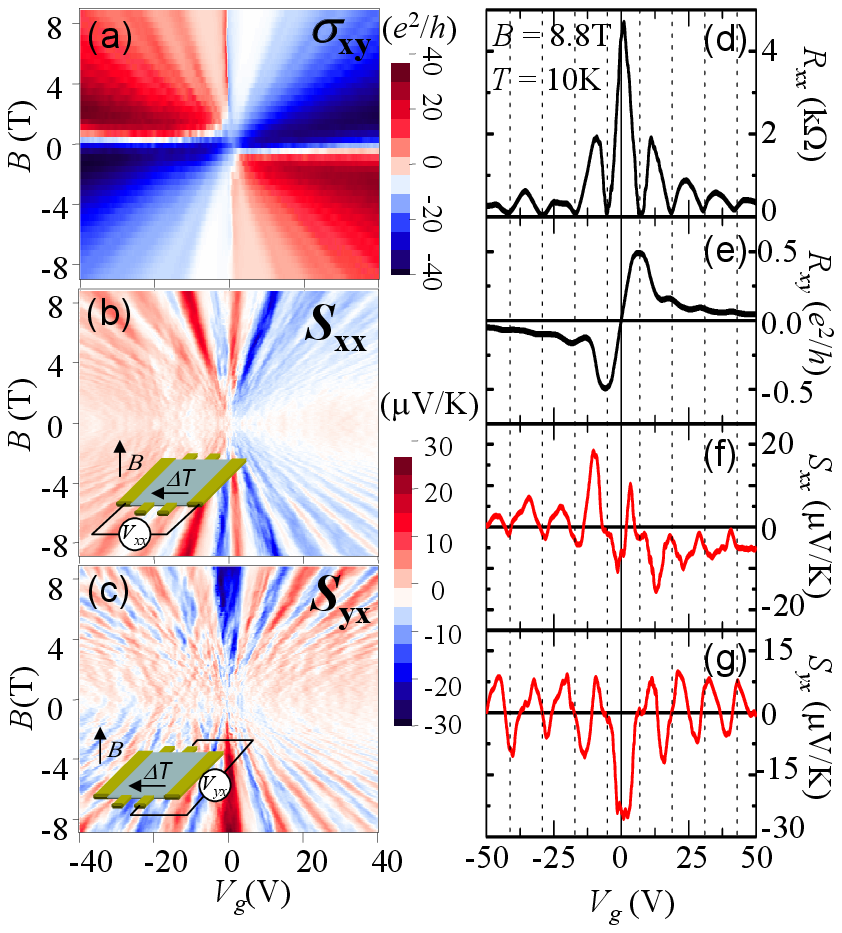}
\caption{ (a) Hall conductance; (b) longitudinal TEP $S_{xx}$; and (c) transverse TEP $S_{yx}$ as a function of $V_g$ and $B$ at $T=10$~K. The insets shows the electrode configurations for $S_{xx}$ and $S_{yx}$ measurement, respectively. (d) $R_{xx}$; (e) $R_{xy}$; (f) $S_{xx}$; and (g) $S_{yx}$ as a function of $V_g$ at a fixed magnetic field $B=8.8$~T and $T=4.2$~K. The vertical doted lines indicate filling factors corresponding to $\nu=\pm2,\pm6,\pm10, \pm14$. All the data is averaged with respect to positive and negative values of magnetic field in order to remove mixing between longitudinal and transverse components.}
\end{figure}

We now turn our attention to the magneto-TEP. In the presence of a perpendicular magnetic field, a Lorentz force bends the trajectories of the thermally diffusing carriers to produce the Nernst TEP, i.e., transverse components of the TEP, $S_{yx}$. For this measurement, we utilize a multiple probe configuration to measure the tensor components of resistivity $R_{xx}$ and $R_{xy}$ and also TEP components $S_{xx}$ and $S_{yx}$ as shown in the inset of Fig. 3(b-c). Fig. 3a displays the Hall conductivity $\sigma_{xy}=-R_{xy}/[R_{xy}^2+(W R_{xx}/L)^2]$ as a function of magnetic field and gate voltage, where $W$ and $L$ are the width and length of the Hall bar device, respectively. The linear growth of the oscillations of $\sigma_{xy}(V_g)$ with increasing magnetic field clearly indicates that Shubnikov-de Hass oscillations (SdH) are present in graphene as previously observed~\cite{geim_QHE,zhang_kim}. Remarkably, similar oscillatory features develop in both $S_{xx}$ and $S_{yx}$ as shown in the Fig. 3(b-c), suggesting that magneto-TEP also provides a sensitive probe for the SdH effect in graphene.

In the high field regime, $B\sim9$~T, fully developed QH effect is observed for filling factors $\nu=\pm2,\pm6,\pm10,\ \pm14$, indicated by the quantized plateaus in $R_{xy}$ and zeros in $R_{xx}$. Vertical dashed lines in Fig.~3(d-g) indicate boundaries between LLs where we find $S_{xx}$ and $S_{yx}$ tend to zero since no carriers are available to participate in diffusion. At the Landau level centers $|S_{xx}|$ is maximal and as higher LL are occupied these peaks decrease. Theoretically, this quantized behavior of $S_{xx}$ is expected in the QH regime of 2D electron gas systems~\cite{streda,jon_girv}. In the low temperature limit, the peaks of $S_{xx}$ are predicted to be given by $S_{xx}^{max}=-\frac{k_B}{e}\frac{\ln{2}}{N+1/2}$ for clean samples, where $N$ in graphene indicates the LL index. Experimentally, thermal and disorder induced broadening of LLs tends to broaden $S_{xx}$, making the quantized values smaller than the above prediction~ \cite{obl_klitz,streda}. The observed $ S_{xx}^{max}$ shows the quantized $1/N$ trend discussed above, but with a reduced factor, indicating the presence of disorder in our graphene samples.

Further analysis which includes the Hall conductivity tensor and the Nernst TEP is performed by employing the generalized Mott formula that holds for all temperatures and can be applied to the QH regime ~\cite{jon_girv,oji}:
 \begin{equation}
S_{ij}^{Mott}=-\frac{\pi^{2}k_{B}^{2}T}{3|e|}\sum_{k}\left({\sigma^{-1}}\right)_{ik}\left(\frac{\partial\sigma}{\partial E_F}\right)_{kj}
 \label{Mott2}
 \end{equation}
where $\sigma_{ij}$ is the conductivity tensor and ${i, j}$ represent the $x$ and $y$ components. According to Eq.~(2), we expect that $S_{xx}/T$ becomes temperature independent in the QH limit where $\sigma_{ij}$ is temperature independent. Indeed, Fig.~4a shows that the measured $S_{xx}/T$ is rather temperature independent for $T< 20$~K. Above this temperature, however, the oscillation amplitude of  $S_{xx}/T$ becomes smaller since thermal activation across the disorder broadened LLs becomes appreciable and the QH effects disappear. Fig.~4b shows good agreement between the measured $S_{xx}$($S_{yx}$) and the calculated $S_{xx}^{Mott}$($S_{yx}^{Mott}$), indicating that the semiclassical Mott relation, extended into the QH regime, works well for both holes ($N<0$) and electrons ($N>0$). Near the zeroth LL ($N=0$) at the CNP ($V_g\approx V_D$), however, the measured $S_{xx}$ exhibits a pair of anomalous oscillation that shows a distinct deviation from Eq. (2). We remark that this pair of $S_{xx}$ peaks near the zeroth Landau level shows opposite TEP polarity, i.e., positive peak on the electron side ($V_g>V_D$) and negative peak on the hole side ($V_g<V_D$). This pair of opposite polarity oscillations could be indicative of a peculiar QH state stemming from the $N=0$ LL. It has been speculated that this LL produces a pair of counter propagating edge states when the Fermi level is at the CNP~\cite{abanin_counter}. In this scenario, the polarity of the inner edge state is opposite to the sign of the bulk majority carrier, which may generate the anomalous TEP observed. In addition, a large enhancement of $S_{yx}$(Inset of Fig.~4b) is observed near the CNP, which also deviates strongly from $S_{yx}^{Mott}$. Such enhancement of the Nernst signal is predicted in conventional 2D electron systems, where the magnitude of $S_{yx}$ strongly depends on the disorder strength~\cite{jon_girv,oji}.

\begin{figure}
\includegraphics[width=1.0\linewidth]{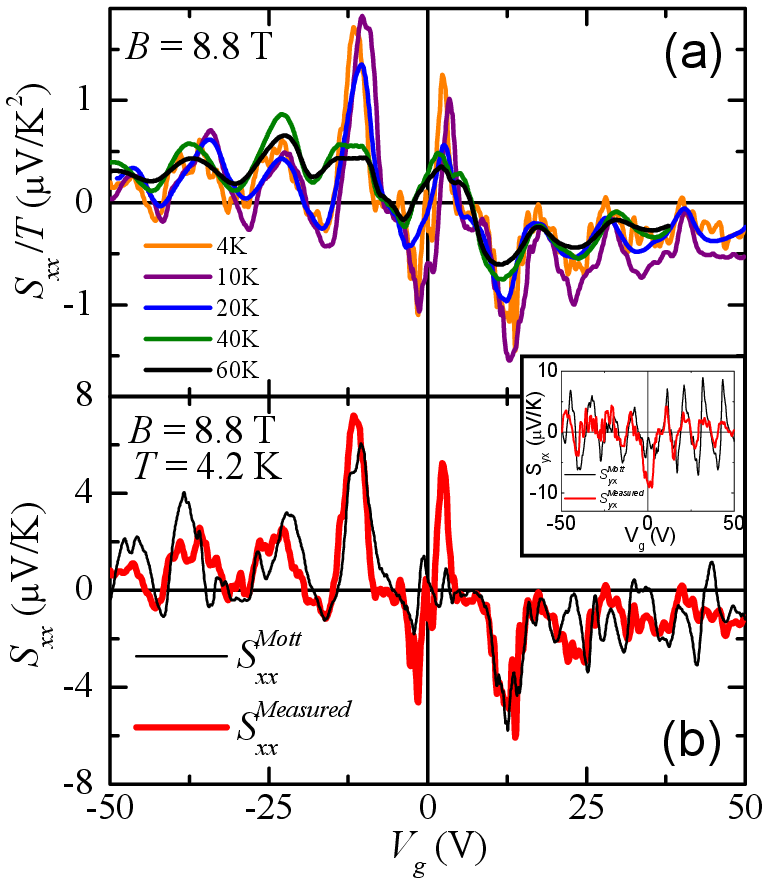}
\caption{(a) Temperature dependence of the $S_{xx}/T$ as a function of $V_g$ at $B=8.8$ ~T. (b) Measured $S_{xx}$ (thick red) and the $S_{xx}$ predicted by the generalized Mott relation (thin black) at $B=8.8$~T and $T=4.2$~K showing diviations near the $N=0$~LL. (Inset) Corresponding $S^{Measured}_{yx}$ and $S^{Mott}_{yx}$.}
\end{figure}

In conclusion, we report carrier density dependent measurements of the thermoelectric coefficient in graphene which scales linearly with temperature. The semiclassical Mott relation explains the observed carrier density modulated TEP particularly well in the low temperature regime. In the QH regime at high magnetic field, the tensor components of the TEP show quantized behavior along with unusual deviations from the generalized Mott relation near the charge neutrality point.

We thank M.S. Foster, I.L. Aleiner, A.F. Young, K.I. Bolotin, and V. V. Deshpande for invaluable discussions. This work is supported by NSF (No. DMR-03-52738), DOE (No. DEFG02-05ER46215), and FENA MARCO grants.

\emph{Note added.}--During the preparation of this manuscript, we became aware of a related work with similar experimental results~\cite{UCR_TEP}.


\begin{thebibliography}{99}
\bibitem{fnote0}
While $\sigma$ provides a bulk conduction quantity without suffering from contact resistance, we found that $G$ has a better match with the measured mesoscopic TEP by the Mott relation which involves a finite number of conduction channels.
\end{thebibliography}
\end{document}